\documentclass[twocolumn,showpacs,preprintnumbers,amsmath,amssymb]{revtex4}

\usepackage{graphicx}
\usepackage{dcolumn}
\usepackage{bm}

\begin{document}

\preprint{APS/123-QED}

\title{Origin of critical-temperature enhancement of an iron-based high-$T_c$ superconductor, LaFeAsO$_{1-x}$F$_{x}$ : NMR study under high pressure
  }

\author{ Tatsuya Nakano$^1$, Naoki Fujiwara$^{1*}$ Kenichiro Tatsumi$^1$,  Hironari Okada$^{2, 3}$, Hiroki Takahashi$^{2, 3}$, Yoichi Kamihara$^{3, 4}$, Masahiro Hirano$^{5, 6}$, and Hideo Hosono$^{4,5, 6}$  }

\affiliation{$^1$ Graduate School of Human and Environmental Studies,
Kyoto University, Yoshida-Nihonmatsu-cyo, Sakyo-ku, Kyoto 606-8501,
Japan}

\email { naoki@fujiwara.h.kyoto-u.ac.jp}

 \affiliation{$^2$ Department of Physics, College of
Humanities and Sciences, Nihon University, Sakurajosui, Setagaya-ku,
Tokyo 156-8550 }

\affiliation {$^3$TRiP, Japan Science and Technology Agency (JST),
Sanban-cho bldg. 5, Sanban-cho, Chiyoda-ku, Tokyo 102-0075, Japan}

\affiliation {$^4$Materials and Structures Labolatory (MSL), Tokyo
Institute of Technology, 4259 Nagatsuda,  Midori-ku, Yokohama
226-8503, Japan}

\affiliation { $^5$ ERATO-SORST, Japan Science and Technology Agency
(JST), Sanban-cho bldg. 5, Sanban-cho, Chiyoda-ku, Tokyo 102-0075,
Japan}

\affiliation{$^6$ Frontier Research Center (FRC), Tokyo Institute of
Technology, 4259 Nagatsuda, Midori-ku, Yokohama 226-8503, Japan }


\date{September 1 2009}

\begin{abstract}
Nuclear magnetic resonance (NMR) measurements of an iron (Fe)-based
superconductor LaFeAsO$_{1-x}$F$_x$ ( $x$ = 0.08 and 0.14) were
performed at ambient pressure and under pressure. The relaxation
rate $1/T_1$ for the overdoped samples ( $x$ = 0.14 )  shows
$T$-linear behavior just above $T_c$, and pressure application
enhances $1/T_1T$ similar to the behavior of $T_c$. This implies
that $1/T_1T = constant$ originates from the Korringa relation, and
an increase in the density of states at the Fermi energy $D(E_F)$
leads to the enhancement of $T_c$. In the underdoped samples ($x$ =
0.08), $1/T_1T$ measured at ambient pressure
 also shows $T$-independent behavior in a wide temperature range above
$T_c$. However, it shows Curie-Weiss-like $T$ dependence at 3.0 GPa
accompanied by a small increase in $T_c$, suggesting that
predominant low-frequency antiferromagnetic fluctuation is not
important for development of superconductivity or remarkable
enhancement of $T_c$. The qualitatively different features between
underdoped and overdoped samples are systematically explained by a
band calculation with hole and electron pockets.
\end{abstract}

\maketitle

LaFeAsO$_{1-x}$F$_x$ is the highly important compound that
stimulated tremendous research activity in Fe-based high-$T_c$
superconductors $^{1)}$. The compound exhibits several phases with F
substitution, i.e., electron doping, on the
temperature-concentration ($T-x$) phase diagram $^{1,2)}$. A
spin-density-wave (SDW)-type antiferromagnetic (AF) ordering of the
parent compound LaFeAsO is suppressed by F substitution, and
superconductivity appears after the AF phase vanishes $^{3-5)}$.
$T_c$ weakly depends on the doping level. The optimal doping level
is around $x$ = 0.11, at which $T_c$ reaches 26 K. Similar phenomena
also appear in a "122" system, (K$_{1-x}$X$_x$)Fe$_2$As$_2$ (X=Sr or
Ba)$^{6-9)}$.

The $T-x$ phase diagram is reminiscent of hole doping in high-$T_c$
cuprates. However, unlike the case of high-$T_c$ cuprates, it is
unclear whether AF spin fluctuation plays an important role in
raising $T_c$. In the case of LaFeAsO$_{1-x}$F$_x$, $T_c$ is
sensitive to pressure ($P$), and shows a clear dome-shaped pressure
dependence on the $T-P$ phase diagram $^{10)}$. The highest $T_c$ is
realized by applying pressure to optimally doped samples ($x \sim $
0.11) or heavily doped samples ($x \sim $ 0.14): $T_c$ of 26 and 20
K for $x$ = 0.11 and 0.14, respectively, goes up to 43 K with
application of a pressure of ~4-5 GPa $^{10-12)}$. (Fig. 1(a).)
However, $T_c$ for lightly doped samples ($x$ = 0.05) hardly goes
beyond 30K even under high pressure $^{11, 12)}$. The suppression of
$T_c$ suggests that a superconducting state with a high $T_c$ is
realized apart from the antiferromagnetically ordered phase on the
$T-x$ phase diagram. To investigate the origin of the $T_c$
enhancement under pressure, and the relationship between the
low-frequency AF spin fluctuation and superconductivity in this
material, we performed $^{75}$As (I = 3/2)-nuclear magnetic
resonance (NMR) measurements under high pressure of underdoped
samples ($x$ = 0.08) and overdoped samples ($x$ = 0.14). To
determine
 $T_c$ and investigate the $P$ dependence, we
also measured resistivity at pressures below 2.6 GPa using a
piston-cylinder-type pressure cell, and resistance above 4 GPa using
a diamond anvil cell.

\begin{figure}
\includegraphics{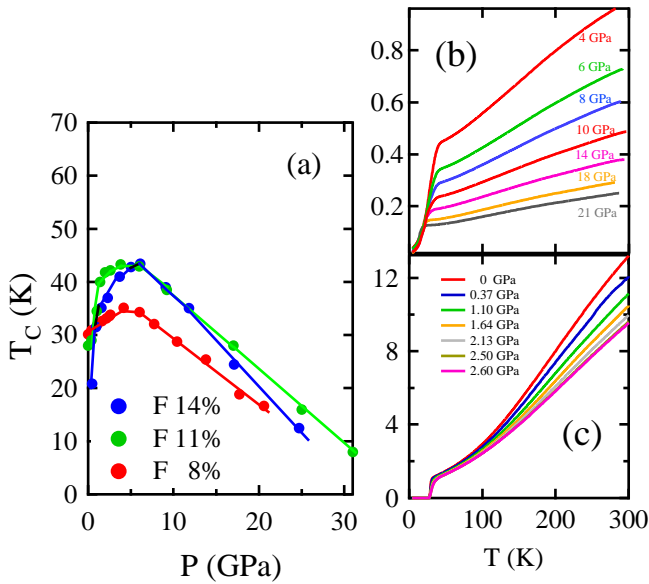}
\caption{\label{fig:epsart}  (a) $P$ dependence of $T_c$ measured
using a piston-cylinder cell and a diamond anvil cell. (b)
Resistance for $8\%$ doped samples at various pressures measured
using a diamond anvil cell. The unit of the vertical axis is
$\Omega$. (c) Resistivity for $8\%$ doped samples at various
pressures measured by using a piston-cylinder cell. The unit of the
vertical axis is $\Omega cm$.}
\end{figure}

The resistance and resistivity for $x$ = 0.08 are plotted in Figs.
1(b) and (c), respectively. The resistivity for $x$ = 0.14 has been
published elsewhere $^{11, 12)}$. Their $T_c$ values are determined
by the onset of superconductivity. In both samples, zero resistivity
was confirmed at low temperatures. The $P$ dependence of $T_c$ for
$x$ = 0.08, 0.11, and 0.14 is plotted in Fig. 1(a). $T_c$ for the
underdoped regime ($x < 0.11 $) does not go beyond  35 K, whereas
that for the overdoped regime ($x > 0.11 $) exceeds 40 K.

We measured NMR spectra under pressure using randomly oriented
powder samples. Field ($\textbf{\emph{H}}$)-swept spectra of the
central transition, $I = -1/2 \Longleftrightarrow 1/2$, show a broad
powder pattern, which prevented accurate Knight shift measurements (
Fig. 2 inset). The line shape is explained by considering the second
order quadrupole effect under a magnetic field $^{13)}$ The
resonance position depends on the angle $\theta$ between
$\textbf{\emph{H}}$ and the maximum electric field gradient (EFG) at
an As nucleus $^{13)}$. The lower- and higher-field peaks in the
inset correspond to $\theta$ = 90 and 42 $^{\circ }$, respectively.
The separation between them is proportional to the square of the
pure quadrupole frequency ($\nu_Q$) except for a minor correction
due to the asymmetry parameter ($\eta$) of the EFG tensor $^{14)}$.
We estimated $\eta$ as 0.1 from the NMR spectra and found that
$\eta$ is insensitive to the doping level. The frequency $\nu_Q$ is
defined as $2\nu_Q = eQV_{zz}/h$, where $Q$ and $V_{zz}$ are the
nuclear quadrupole moment and maximum EFG, respectively. The
separation between the peaks decreases with increasing pressure. The
results of $\nu_Q$ at several pressures are plotted in the main
panel of Fig. 2. The frequency $\nu_Q$, i.e., $V_{zz}$ originates
from the on-site charge density and the surrounding Fe ions.
$V_{zz}$ is sensitive to the distance between Fe and As ions: a
stretching of the Fe-As distance decreases EFG originating from the
surrounding Fe ions and weakens the hybridization between As-4p and
Fe-3d orbitals, which would lead to the decrease in the on-site
charge density. The decrease in $\nu_Q$ or EFG due to application of
pressure can be explained by the stretching of the Fe-As distance.
The stretching due to application of pressure has been observed from
synchrotron radiation measurements under high pressure $^{15)}$.

\begin{figure}
\includegraphics{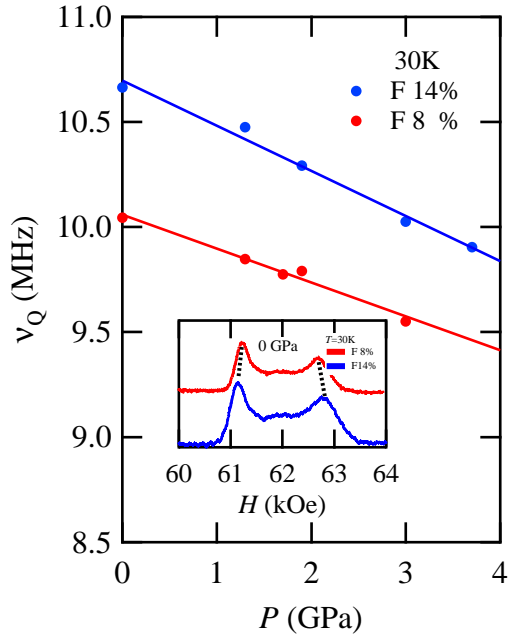}
\caption{\label{fig:epsart} $P$ dependence of pure quadrupole
frequency of $^{75}$As nuclei. The inset shows field-swept spectra
of the central transition, $I = -1/2 \Longleftrightarrow 1/2$. }
\end{figure}

We measured the relaxation rate $1/T_1$ at $\theta$ = 90 $^{\circ}$
using a saturation recovery method. The $T$ dependence of  $1/T_1$
for $x$ = 0.08 and 0.14 is shown in Figs. 3(a) and (b),
respectively. The $T$ dependence of $1/T_1T$ is shown in Fig. 4.
$1/T_1$ for the two doping levels shows qualitatively different $T$
dependence.

\begin{figure}
\includegraphics{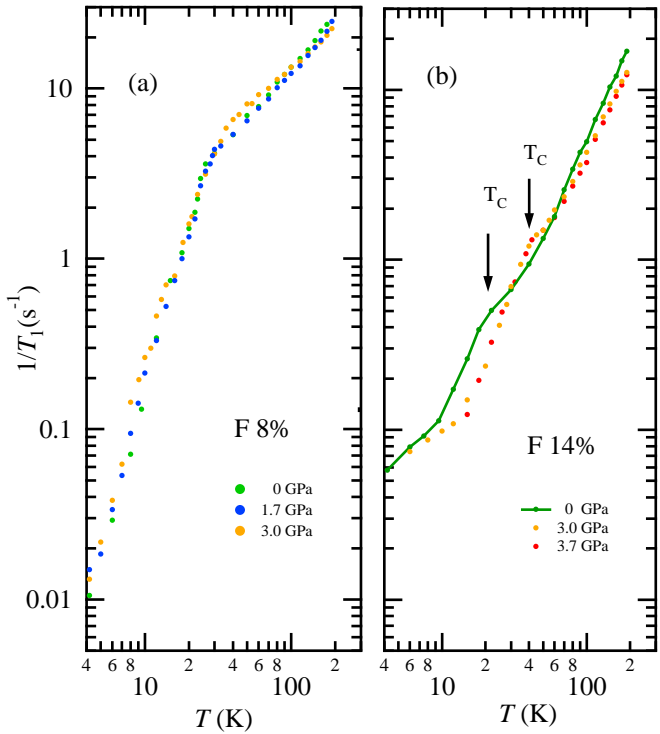}
\caption{\label{fig:epsart} $^{75}$As-nuclear magnetic relaxation
rate $1/T_1$ for $\textbf{H} \bot$ the maximum electric field
gradient of $^{75}$As. (a) Underdoped regime ($x$ = 0.08). (b)
Overdoped regime($x$ = 0.14).
 }
\end{figure}

\begin{figure}
\includegraphics{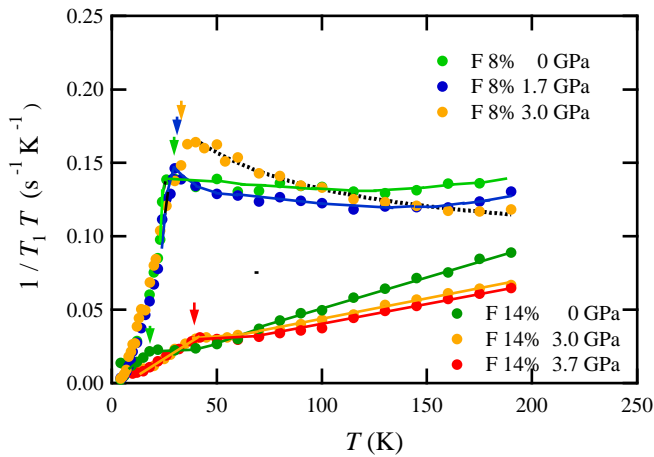}
\caption{\label{fig:epsart}$1/T_1T$ measured at several pressures.
 The dotted curve represents a
Curie-Weiss curve. The other lines are guides to the eye.  }
\end{figure}

In the case of $x$ = 0.14, $T$-linear dependence is observed in a
narrow $T$ range just above $T_c$. The $T$-linear dependence is
clearly observed as a plateau in Fig. 4. $T_c$ values determined
from the resistivity measurements are indicated by arrows in Figs. 3
(b) and 4. The value of $1/T_1T$  is enhanced with increasing
pressure similar to the behavior of $T_c$. $1/T_1T$ seems to change
in accordance with $T_c$ with increasing pressure: both $1/T_1T$ and
$T_c$ increase remarkably with increasing pressure from 0 to 3.0
GPa, and the change in $1/T_1T$ between 3.0 and 3.7 GPa is small,
similar to that in $T_c$. The $T$-linear dependence is attributable
to the Korringa relation, and the value of $1/T_1T$ just above $T_c$
is proportional to the square of the density of states (DOS) at the
Fermi energy, $D(E_F)$. At high temperatures, deviation from the
$T$-linear dependence becomes remarkable. The increase in $1/T_1T$
can be explained by a characteristic band structure of this system,
as described below $^{16)}$. At low temperatures, another $T$-linear
dependence appears, suggesting the existence of the impurity
scatterings $^{17-19)}$. In the overdoped regime, the system can be
well described as a band metal, and application of pressure causes
an increase in $D(E_F)$ and enhancement of $T_c$.

In the case of $x$ = 0.08, $1/T_1$ shows $T$-linear dependence in a
wide $T$ range above $T_c$ at ambient pressure, and $T^3$ dependence
below $T_c$, as already reported by other groups $^{20-22)}$. $T_c$
values determined from the resistivity measurements, indicated by
arrows in Fig. 4, are consistent with those estimated from the
change in $1/T_1T$ within an accuracy of several Kelvins. At first
glance, the $T$-linear dependence is reminiscent of the Korringa
relation, as in the case of $x$ = 0.14. However, it does not
originate from the conventional Korringa relation: if the $T$-linear
dependence originates from the Korringa relation, the estimated
$T_c$ should go beyond 40 K because the value of $1/T_1T$, namely,
$D(E_F)$ for $x$ = 0.08 is much larger than that for $x$ = 0.14.
Furthermore, $1/T_1T \sim constant$, observed at ambient pressure,
breaks down under high pressure, as seen from the data at 3.0 GPa.
$1/T_1T$ increases monotonously toward $T_c$. $1/T_1T$ at 1.7GPa
shows transitional behavior from $1/T_1T \sim constant$ to
Curie-Weiss behavior. The dotted curve in Fig. 4 represents a
Curie-Weiss curve: $1/T_1T = 0.09 + 6.2/(T + 39)$ $ (s^{-1}K^{-1})$.
The Curie-Weiss behavior is reminiscent of high-$T_c$ cuprates.
Although low-frequency AF fluctuation predominates by applying
pressure, an increase in $T_c$ is small. Low-frequency AF
fluctuation is not essential to achieving the highest $T_c$,
although it would contribute to raising $T_c$ to some extent: the
highest $T_c$ is realized for $x$ = 0.11-0.14 without development of
low-frequency AF fluctuation. It is concluded that low-frequency AF
fluctuation suppresses the development of superconductivity or the
enhancement of $T_c$ in this material.

The qualitatively different features between samples with $x$ =0.08
and 0.14 are explained by a scenario based on a band calculation
with electron and hole pockets $^{23, 24)}$. The system can be
treated as a simple two-dimensional square lattice of an Fe atom,
although two Fe atoms are contained in the actual unit cell. In the
unfolded Brillouin zone (Bz), hole pockets exist around $\Gamma (0,
0)$ and $\Gamma' (\pi, \pi)$ in addition to electron pockets around
M points $^{23)}$.  $\Gamma' (\pi, \pi)$ overlaps $\Gamma (0, 0)$ in
the original folded Brillouin zone. With increasing doping level,
the Fermi energy moves upward, and the hole pockets around $\Gamma
(\Gamma') $ become smaller. The hole pocket around $\Gamma' (\pi,
\pi)$ is sensitive to the doping level, and it first vanishes with
increasing doping level, as illustrated in Fig. 5 $^{25}$. In the
underdoped regime, the nesting between electron and hole pockets
gives rise to AF fluctuation, which predominates and would suppress
development of superconductivity, namely remarkable enhancement of
$T_c$. Application of pressure seems to promote the nesting. In the
overdoped regime, the nesting becomes weak when electron doping
moves the Fermi energy upward and the hole pockets around $\Gamma
(\Gamma')$ become smaller. In such a situation, remarkable
enhancement of $T_c$ is possible. When the hole pocket around
$\Gamma'$ vanishes, a large weight of DOS still remains just below
the Fermi energy as illustrated in Fig. 5. The contribution from
these energy levels presumably leads to an increase in $1/T_1T$ at
high temperatures, as seen in the $x =0.14$ doped samples. The
scenario can be expanded to other systems, such as hole-doped
systems: in this case the Fermi energy moves downward with
increasing doping level, and AF fluctuation predominates until the
Fermi energy comes across the bottom of the electron band, which
would suppress any rise in $T_c$. This scenario may answer the
question of why the highest $T_c$ (over 50 K) is realized only for
"1111" systems. To investigate the origin of the highest $T_c$
observed in a "1111" system, the pressure effect on the electron
pockets around M seems important.

\begin{figure}
\includegraphics{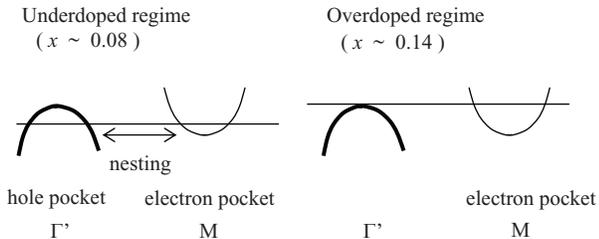}
\caption{\label{fig:epsart}  A scheme of a two-band model. In the
underdoped regime, nesting between hole and electron pockets causes
antiferromagnetic fluctuation as seen in $1/T_1T$ for the 8\% doped
samples at 3.0 GPa ( Fig. 4). In the overdoped regime, the hole
pockets around $\Gamma'$ disappears, and the system can be described
as a band metal. The behavior is seen in $1/T_1T$ for the 14\% doped
samples (Fig. 4). }
\end{figure}

We would like to thank K. Kuroki, and H. Ikeda for fruitful
discussions and A. Hisada for experimental support. This work was
partially supported by a Grant-in-Aid (KAKENHI 17340107) from the
Ministry of Education, Science and Culture, Japan.


\ \\






  \ \\

\end{document}